\documentstyle[11pt,newpasp,twoside,epsf]{article}
\begin{document}

\title{VLBI Techniques in Pulsar Astronomy}
\author{Walter Brisken}
\affil{National Radio Astronomy Observatory, PO Box O, Socorro, NM 87801, USA}

\begin{abstract}
Very long baseline interferometry (VLBI) provides the resolution needed to
make precision measurements of a pulsar's parallax and proper motion.  In
making these measurements, the astronomer is faced with difficult 
calibration problems and a paucity of strong point-like calibrators
that ultimately limit the accuracy of parallax
measurements of even the brightest pulsars.  A new technique to 
calibrate away the effects of the ionosphere, the dominant source of
phase error at frequencies below 5~GHz, has led to the new measurements of
nine pulsar parallaxes to 100~microarcsecond or better precision.
\end{abstract}

\section{Introduction}

The case for measuring pulsar parallaxes is strong -- many models based on
other measurements of pulsars require estimates of the distances to the
pulsars in the sample.  Applications of pulsar distances are not within the
scope of this paper as they are discussed in detail by Chatterjee in the 
following paper.  This paper concentrates on making useful measurements of
pulsars with VLBI and focuses mainly on high-precision astrometry.

\section{VLBI}

Interferometry is the radio astronomers answer to the optical astronomer's 
resolution.  
Building large single centimeter-wave telescopes with resolutions that 
come close to those of large optical telescopes is impossible.  At 1.5~GHz, the
dish would need to be 20~km in diameter -- quite a mechanical feat.  Even if
such a marvel could be constructed, the ionosphere and troposphere could not
be expected to be coherent over the entirety of the aperture.  Fortunately
phase-preserving amplifiers and stable electronics are available in the
gigaHertz frequencies which allow the electrical combination 
of signals from many small antennas.  Also important is that the radio
photons are in the classical regime with occupation numbers much greater
than unity, which allows amplification and duplication of the signals without
a large sensitivity penalty.  Once one can combine the signals from many small
antennas, one can use aperture synthesis to probe features smaller than
a milliarcsecond.

VLBI is probably best distinguished from connected-element interferometry
by its scale -- thousands of kilometers, typically.  
The distance and geographic barriers require the sampled electric field 
measurements to be recorded onto (usually magnetic) media to be shipped
from the antennas to a central correlator and the use different clocks at each 
station which must be stable and synchronized to better than microsecond 
precision.  It also requires use of very detailed correlator models,
especially for astrometric projects.  These models include effects such
as bending of light around the sun and Jupiter, tidal deformation of the 
earth, excursions from Earth's average spin, and plate tectonic motion.  
This model is used to calculate, to 10~ps or better accuracy, the expected
geometric delay between the arrivals of the radiation from an astronomical 
source at the antennas. 
The input parameters to the model include station positions, the 
time of the observation, and coordinates of the source being observed (the
phase center).
The data products of the correlator are the cross-correlations of
the signals from each pair of stations after taking into account the
modeled delay.  These are called visibilities.  

\section{Scattering}

This paper will not focus on scattering, however for completeness I include
a brief discussion.  VLBI with its high angular resolution can measure
the scattering disc size, constraining the distribution of scattering
material along the line of sight to the source.  Scattering increases
strongly (as $\lambda^2$) with wavelength.  

While an image that exhibits the scattering disc can be
reconstructed with the set of visibilities measured for a source, a more
accurate measurement can be made by modeling the visibilities directly.
The amplitude of a
visibility is the flux density of the source being observed multiplied by
the correlation coefficient, which for a point source would be unity.
Earth rotation causes the projected antenna separations (baseline lengths)
to change, probing a wide range of baselines.  A detection of a scattering
disc manifests itself as a drop in the correlation coefficient as a function
of baseline length.
The scattering-based decorrelation phenomenon can
be useful for probing the ISM, but is generally a nuisance for VLBI at
low frequencies as it reduces the effective resolution.

\section{VLBI astrometry}

Astrometry is the science of making precision measurements of positions of
astronomical sources.  Multi-epoch astrometry leads to
measurements of proper motions and perhaps parallaxes.  
The phases of the complex visibilities contain astrometric information.  In
the absence of correlator model errors or other deviations from perfect
calibration, the visibility phase for a point source is given by
\begin{equation}
\phi(\nu) = \frac{\nu}{c}\left(u l + v m\right),
\end{equation}
where $c$ is the speed of light, $\nu$ is the observing frequency,
$(u, v)$ is the projected baseline vector, and $(l, m)$ is the displacement
vector of the source relative to the correlator model phase center.  
For a source that is exactly at the correlator phase center,
this phase is zero.  With multiple measurements of $\phi$ made with different
projected baselines, $(u, v)$, the source position, $(l, m)$, can be
obtained.

Pulsar VLBI measurements are made almost exclusively in the 20~cm observing
band.  Pulsars have steep spectral indexes, making observing at higher
frequencies more difficult.  The ionosphere, scattering, interference,
and lower angular resolution all contrive to make observing at frequencies
below 1~GHz difficult.  Pulsar gating is often employed to improve the
signal-to-noise of the pulsar measurements.  Gating disables accumulation in
the correlator during a pulsar's off-pulse phase.

\subsection{Phase-referencing} \label{sec:PhaseRef}

The ionosphere,
just like the ISM, contains a non-homogeneous distribution of electrons which
impart a dispersive delay to the incoming radiation.  The ionosphere
above each station can change substantially on timescales less than 
1~minute, causing the relative phases to change, corrupting the astrometry.
The bulk of this can be removed by a technique known as phase-referencing
(Shapiro et al. 1979).
In this process, a nearby point-like calibrator source with a well known
position is observed alternately
with the target with a cycle time of 2 to 4 minutes.  Since the calibrator's
position is well known, its visibility phases can be assumed to be purely
due to calibration problems.  These phases are then interpolated across the
pulsar's observation and subtracted from the pulsar's corresponding visibility 
phases.  This standard observing technique calibrates away most of the
ionosphere's effect, but the uncorrected ionosphere gradients often leave
significant and systematic phase errors.   

\subsection{Ionospheric calibration} \label{sec:Iono}

After phase-referencing at frequencies lower than
about 5~GHz, the remaining phase errors are dominated by the effects of
the ionosphere.
The dispersive nature of the ionosphere can be exploited in its removal.
The unwanted phase as a function of frequency for a given visibility can
be expressed as
\begin{equation}
\phi^{\mathrm{iono}}(\nu) = \frac{B}{\nu},
\end{equation}
where B is the strength of the uncalibrated ionosphere.
The phase-referenced visibility phase then has the form
\begin{equation}
\phi^{\mathrm{vis}}(\nu) = \frac{\nu}{c}\left(u l + v m\right) + \frac{B}{\nu}.
\end{equation}
The difference in frequency dependence of the first, wanted term and the
second, unwanted term make it possible to measure $B$ and remove its
effect from the visibility phase, allowing precise astrometry.
Additional details of this
technique and a discussion of its applicability can be found in 
Brisken et al. (2000 \& 2002).
A project to measure the parallaxes of 10 pulsars was started in 1999 after
success of this technique was demonstrated on pulsar B0950+08 
(Brisken et al. 2000).  
In this
experiment, the NRAO\footnote{The National Radio Astronomy Observatory is a facility of the National Science Foundation operated under cooperative agreement by Associated Universities, Inc.} Very Long Baseline Array (VLBA) was used in the 
1400--1730~MHz band.  Eight frequency sub-bands, each 8 MHz wide, were placed
across 330~MHz.  Five epochs of observation were planned for each pulsar over
the course of one year.  One pulsar, J2145-0750, was not detected in either
of its first two epochs and was subsequently abandoned.  Successful
parallax measurements were made for the nine remaining pulsars with a typical
precision of 100 microarcseconds (Brisken et al. 2002).  
See Table~\ref{tab:parallaxes} for the results.

\begin{figure}
\label{fig:0950}
\plottwo{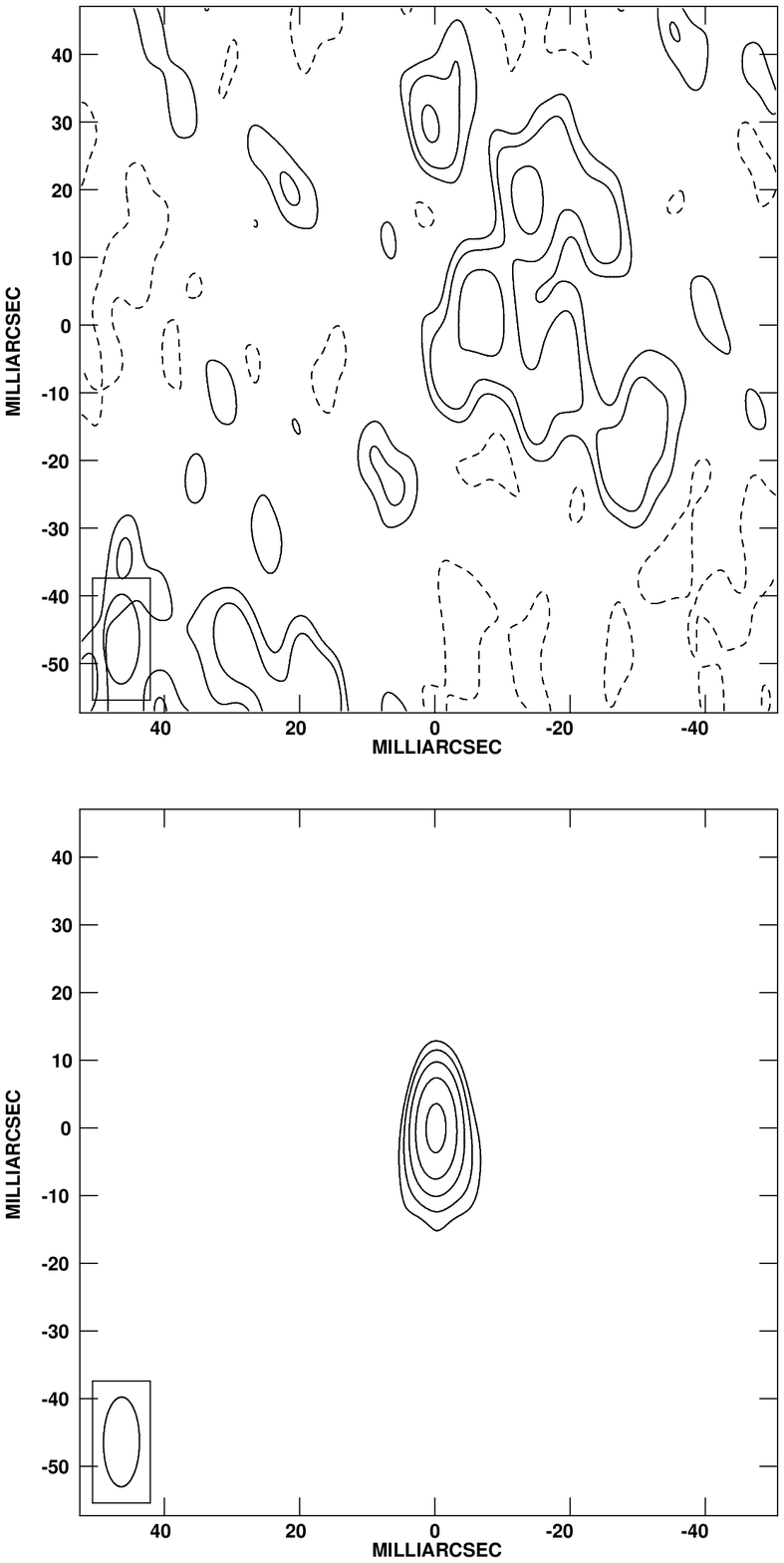}{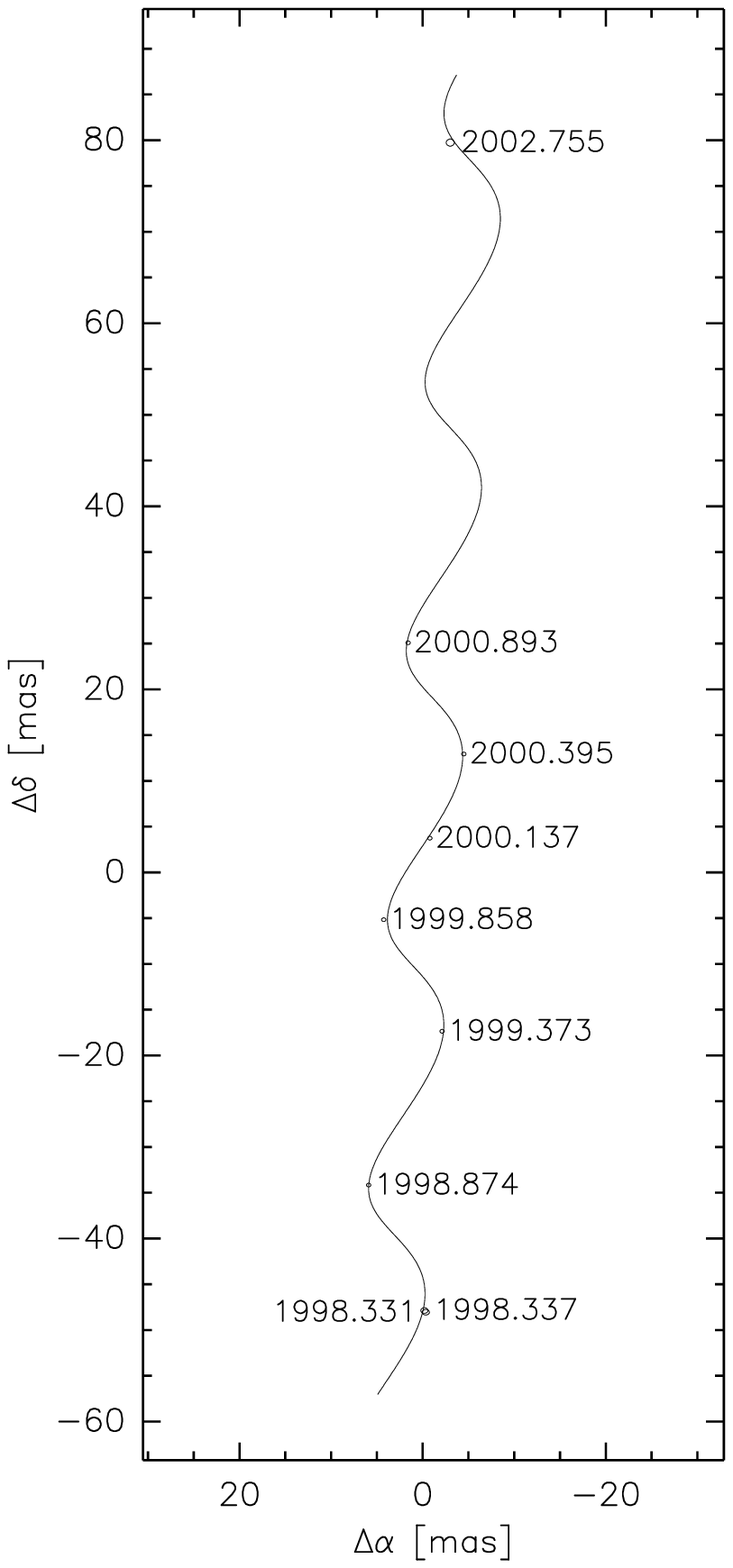}
\caption{Ionospheric calibration applied to B0950+08.  The top left figure
is a contour plot showing the image made of B0950+08 prior to ionosphere
calibration.  The bottom left figure shows the same after ionosphere
calibration.  The lowest contour in each image is at 5~mJy beam$^{-1}$.  
A factor of 2
separates consecutive contour levels.  The beam size for each is 4 by 11
milliarcseconds.  The right figure shows the best fit parallax 
($\pi = 3.81 \pm 0.07$~mas) and proper motion 
($\mu_\alpha = -2.06 \pm 0.06$~mas~yr$^{-1}$, 
 $\mu_\delta = 29.37 \pm 0.05$~mas~yr$^{-1}$).
The data point from year 2002.755 used only 3 VLBA stations; the Mark~5
recording system was used at the Pie Town station.  Note that the sizes of the
ellipses near each date represent the positional uncertainty at that epoch.
}
\end{figure}

\subsection{In-beam calibration} \label{sec:InBeam}

A second method to improve phase-referencing is to use a phase-reference
calibrator source that is very close to the target.  In the case that a
compact source of sufficient flux density (about 5~mJy for the VLBA) is
within the same primary beam as the target (about 25~arcminutes for the
VLBA at $\lambda \sim 20$~cm), both can be observed simultaneously.  At
correlation time, each source is correlated separately with its own 
phase center.  The simultaneous observation of the calibrator and target
mean that no temporal interpolation is needed and the spatial proximity
implies that ionospheric gradients contribute little to the phase-referenced
visibility phase.  Finding a suitable in-beam calibrator is often difficult
or impossible.  Once a good calibrator is found, a successful parallax 
measurement is almost guaranteed.

\begin{table}
\caption{Parallaxes of radio pulsars}
\label{tab:parallaxes}
\begin{tabular}{lclc}
\tableline
      Pulsar
    & $\pi$
    & Technique$^\dagger$
    & Reference
\\
    & (mas)
    &
    &
\\
\tableline
B0329+54     & $0.94 \pm 0.11$ & Iono$^1$      & (1) \\
J0437$-$4715 & $7.19 \pm 0.14$ & Timing        & (2) \\
B0809+74     & $2.31 \pm 0.04$ & Iono          & (1) \\
B0823+26     & $2.8  \pm 0.6 $ & Phase-Ref$^2$ & (3) \\
B0833$-$45   & $3.4  \pm 0.7 $ & Optical (HST) & (4) \\ 
B0919+06     & $0.83 \pm 0.13$ & In-beam$^3$   & (5) \\
B0950+08     & $3.81 \pm 0.07$ & Iono          & (1) \\
B1133+16     & $2.80 \pm 0.16$ & Iono          & (1) \\
B1237+25     & $1.16 \pm 0.08$ & Iono          & (1) \\
B1451$-$68   & $2.2  \pm 0.3 $ & In-beam       & (6) \\
B1534+12     & $0.91 \pm 0.13$ & Timing        & (7) \\
J1713+0747   & $0.9  \pm 0.3 $ & Timing        & (8) \\
J1744$-$1134 & $2.8  \pm 0.2 $ & Timing        & (9) \\
B1855+09     & $1.1  \pm 0.3 $ & Timing        & (10) \\
B1929+10     & $3.02 \pm 0.09$ & Iono          & (1) \\
B2016+28     & $1.03 \pm 0.10$ & Iono          & (1) \\
B2020+28     & $0.37 \pm 0.12$ & Iono          & (1) \\
B2021+51     & $0.50 \pm 0.07$ & Iono          & (1) \\
\tableline
\tableline
\end{tabular}

$\dagger$ 1: see Sec.~\ref{sec:Iono}; 2: see Sec.~\ref{sec:PhaseRef};
3: see Sec.~\ref{sec:InBeam}
References:
(1) Brisken et al., 2002;
(2) van Straten et al. 2001;
(3) Gwinn et al., 1986;
(4) Caraveo el al., 2001;
(5) Chatterjee et al., 2001;
(6) Bailes et al., 1990;
(7) Stairs et al., 1999;
(8) Camilo et al., 1994;
(9) Toscano et al., 1999;
(10) Kaspi et al., 1994.
\end{table}

\subsection{Disc-based recording}

VLBI has traditionally relied on large (in physical dimension and 
data capacity) magnetic tapes for storage and
transportation of data from the antennas to the correlator.  Market pressure
over the last decade has forced the price of commodity IDE discs to fall
below that of magnetic tape.  Discs have additional benefits over tapes for 
reasons including: random access to data is possible,
much less maintenance is required, recording rates can be
greater and capacity scales with the consumer market.  The pulsar community
may greatly benefit from the deployment of disc-based VLBI recorders at most
large radio telescopes in the near future. 

Haystack Observatory is leading the Mark~5 disc-based recorder project.
To date, roughly 20 prototype units are deployed.  The VLBA achieved first
fringes with its first Mark~5 prototype on 13~Oct.~2002.  Pulsar B0950+08
and its phase reference calibrator were observed with magnetic tape at the
St.~Croix and Hancock, NH VLBA stations and onto a Mark~5 recorder at the 
Pie Town, NM VLBA station.  This data point augments 8 previous epochs and is
shown in Fig.~\ref{fig:0950}.  The updated parallax
for B0950+08 is shown in Table~\ref{tab:parallaxes}.

\end{document}